\documentclass[twocolumn,
 amsmath,amssymb,
 aps,
 pra,
]{revtex4}

\usepackage{graphicx}
\usepackage{dcolumn}
\usepackage{bm}
\usepackage[colorlinks,urlcolor=blue,citecolor=blue,linkcolor=blue]{hyperref}
\usepackage{comment}
\usepackage{color}

\renewcommand{\k}{{\bf k}}
\newcommand{\p}{{\bf p}}
\newcommand{\q}{{\bf q}}

\newcommand{\0}{{\bf 0}}

\newcommand{\ab}{a_{\rm B}}

\newcommand{\ket}[1]{\left|{#1}\right>}

\newcommand{\ek}{\epsilon_{\k}}
\newcommand{\ep}{\epsilon_{\p}}

\newcommand{\nn}{\nonumber}
\newcommand{\beq}{\begin{equation}}
\newcommand{\eeq}{\end{equation}}

\newcommand{\B}{\textnormal{B}}

\newcommand{\T}{{\cal T}}

\newcommand{\tb}{{\cal T}_{\text{B}}}
\newcommand{\mb}{m_{\text{B}}}
\renewcommand{\ab}{a_{\text{B}}}

\newcommand{\sout}[1]{}
\begin{document}

\title{Finite-temperature behavior of the Bose polaron}

\author{Jesper Levinsen}
\affiliation{School of Physics and Astronomy, Monash University, Victoria 3800, Australia}
\author{Meera M.~Parish}
\affiliation{School of Physics and Astronomy, Monash University, Victoria 3800, Australia}
\author{Rasmus S.~Christensen}
\affiliation{Department of Physics and Astronomy, Aarhus University, DK-8000 Aarhus C, Denmark.}
\author{Jan J. Arlt}
\affiliation{Department of Physics and Astronomy, Aarhus University, DK-8000 Aarhus C, Denmark.}
\author{Georg M.~Bruun}
\affiliation{Department of Physics and Astronomy, Aarhus University, DK-8000 Aarhus C, Denmark.}

\date{\today}

\begin{abstract}
We consider a mobile impurity immersed in a Bose gas at finite temperature.
Using perturbation theory valid for weak coupling between the impurity and the bosons, we derive analytical results for the 
energy and damping of the impurity for low and high temperatures, as well as 
for temperatures close to the critical temperature $T_c$ for Bose-Einstein condensation. 
These results show that the properties of the impurity vary strongly with temperature. In particular, the energy exhibits a non-monotonic behavior close to $T_c$, and the damping rises sharply  close to $T_c$. We argue that this behaviour is generic for impurities immersed in an environment undergoing a phase transition that breaks a continuous symmetry. 
Finally, we discuss how these effects can be detected experimentally. 
\end{abstract}

\maketitle

\section{Introduction}

The experimental realization of highly population-imbalanced atomic gases has 
dramatically improved our understanding of the properties of mobile impurities in a quantum 
medium. Using Feshbach resonances~\cite{Chin2010} to tune the  interaction 
between the impurity and the reservoir, cold-atom experiments have systematically explored the properties 
of impurities first in fermionic~\cite{Schirotzek2009,Kohstall2012,Koschorreck2012} and recently also in 
bosonic~\cite{Jorgensen2016,Hu2016} reservoirs. While there are many similarities between impurities 
in fermionic and bosonic reservoirs (termed the Fermi and Bose polaron, respectively), there are also 
important differences. For instance, whereas the Fermi polaron has a sharp transition to a molecular state with increasing attraction~\cite{Chevy2006,Prokofev2008,Mora2009,Punk2009,Combescot2009,Cui2010,Massignan2011,Massignan_Zaccanti_Bruun,Cui2015}, the Bose polaron exhibits a smooth crossover 
instead, either to a molecular state~\cite{Rath2013} or the lowest Efimov trimer~\cite{Levinsen2015} depending on the value of the three-body parameter. The Bose polaron has also been proposed to be unstable towards 
other lower lying states~\cite{Shchadilova2016,Grusdt2017}. 

Here, we investigate a unique feature of the Bose polaron (polaron from now on): 
The medium  exhibits a phase transition between a Bose-Einstein condensate (BEC) and a normal gas.
The effect of such a transition on the quasiparticle properties has not been explored before in previous finite-temperature studies of the Bose polaron~\cite{Boudjema2014,Schmidt2016}.
Using perturbation theory valid for
weak coupling, we show  that this transition gives rise to several interesting effects. Both the energy and the damping of the polaron depend strongly and 
in a non-trivial way on the temperature in the region around the critical temperature $T_c$. 
More generally, these effects are relevant to the behavior of quasiparticles near a phase transition that breaks a continuous symmetry of the system.
We discuss how these  effects can be 
measured. Very recently, the temperature dependence of the polaron was investigated for strong coupling~\cite{Guenther2017}. Our present study focuses instead on the weak-coupling regime where rigorous results can be derived. 

The paper is organized as follows. In Sec.~\ref{sec:model} we describe the model and introduce the perturbative framework. Our main results are presented in Sec.~\ref{sec:results}. Here we describe the polaron properties in three different temperature regimes: at low temperature, in the region close to the critical temperature for Bose-Einstein condensation, and all the way to high temperature. We conclude in Sec.~\ref{sec:conc}.

\section{Model and methods}
\label{sec:model}
We consider an impurity of mass $m$ in a gas of bosons with
mass $m_{\B}$. The Hamiltonian is 

\begin{align}
   H=&\sum_\k\ek^{\B \vphantom{\dagger}}b^\dagger_\k b_\k^{\vphantom{\dagger}}
   			+ \frac{g_\B}{2} \sum_{\k,\k',\q}
 b^\dagger_{{\k}+{\q}}b_{{\k}'-\q}^\dagger
 b_{{\k}'}^{\vphantom{\dagger}} b_{{\k}}^{\vphantom{\dagger}}
\nonumber\\
& +\sum_{{\k}}\epsilon_{\k}^{\vphantom{\dagger}} c^\dagger_{{\k}}c_{{\k}}^{\vphantom{\dagger}}
 +g \sum_{{\k},{\k}',{\q}}c_{{\k}+\q}^\dagger b^\dagger_{{\k}'-{\q}} b_{{\k}'}^{\vphantom{\dagger}} c_{{\k}}^{\vphantom{\dagger}}, 
\end{align}
where the operators $b^\dag_{{\k}}$ and $c^\dag_{{\k}}$ create a boson and the impurity, respectively, with momentum ${\k}$ and free dispersions
$\ek^{\B}=k^2/2m_{\B}$ and $\ek={k^2}/{2m}$. The boson-boson and the boson-impurity interactions are short range with coupling strengths $g_\B$ and $g$, respectively, and 
we work in units where the volume, $\hbar$, and $k_B$ are  1.

The Bose gas is taken to be weakly interacting, i.e.,
$n\ab^3\ll1$, where $n$ is the boson density and $\ab>0$ is the 
boson-boson scattering length. As we are interested in deriving rigorous results, we use Popov theory to describe the Bose gas. 
Below the BEC critical temperature 
$T_c\simeq \frac{2\pi}{[\zeta(3/2)]^{2/3}}\frac{n^{2/3}}{m_\B}$, we have the usual Bogoliubov dispersion
 $E_{\mathbf k}=[\ek^\B(\ek^\B+2\tb n_0)]^{1/2}$, where
$n_0$ is the condensate density, and $\tb=4\pi \ab/\mb$ the boson vacuum scattering matrix.
Below $T_c$, we have the normal and anomalous propagators for the 
bosons in the BEC, 
\begin{align}
G_{11}(\mathbf k,i\omega_s)&=\frac{u^2_\mathbf k}{i\omega_s-E_{\mathbf k}}-\frac{v^2_{\mathbf k}}{i\omega_s+E_{\mathbf k}}\nonumber\\
G_{12}({\mathbf k},i\omega_s)=G_{21}({\mathbf k},i\omega_s)
&=\frac{u_{\mathbf k} v_{\mathbf k}}{i\omega_s+E_{\mathbf k}}-\frac{u_{\mathbf k} v_{\mathbf k}}{i\omega_s-E_{\mathbf k}}.
\label{BoseGreens}
\end{align}
where $u_{\mathbf k}^2=1+v_{\mathbf k}^2=[(\ek^\B+\tb n_0)/E_{\mathbf k}+1]/2$ are the coherence factors, and $\omega_s=i2sT$ is a boson Matsubara frequency with $s$ integer. 
The condensate density 
is then found self-consistently from the condition 
\begin{align}
n&=n_0-T
\sum_{\omega_s,\k}e^{i\omega_s0_+} 
G_{11}(\mathbf k,i\omega_s)\nn\\
& = n_0+\frac{8n_0}{3\sqrt\pi}(n_0a_B^3)^{1/2}+
\sum_\k\frac{\epsilon_\k^{\rm{B}} + \tb n_0}{E_\k} f_\k.
\label{eq:npopov}
\end{align}
where  
$f_\k=[\exp(E_\k/T)-1]^{-1}$ is the Bose distribution function for temperatures $T<T_c$.
Popov theory  
provides an accurate description
except in a narrow critical region 
determined by $|T-T_c|/T_c\lesssim n^{1/3}\ab$~\cite{Shi1998}.

\begin{figure}[tbp]
\centering
\includegraphics[width=\columnwidth]{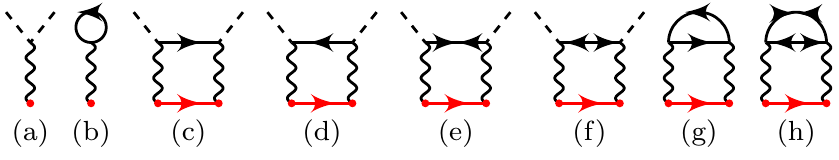}
\caption{(a-b) First and (c-h) second order diagrams for the impurity self-energy. The impurity propagator is shown as the bottom red lines, and the external impurity propagators attach to the red dots. The boson normal and anomalous
propagators are shown as the upper solid black lines, while dashed lines are condensed bosons.
The wavy vertical lines denote the impurity-boson 
  scattering matrix $\T_v$.}
\label{fig:diagrams}
\end{figure}

\subsection{Perturbation theory} 

We use perturbation theory in powers of the impurity-boson scattering length $a$ to analyze the  impurity problem. At $T=0$, this approach has  yielded important information. For instance, 
the impurity energy was shown to depend logarithmically on $a$ at third order~\cite{Christensen2015},
similarly to the energy  of a weakly interacting Bose gas beyond Lee, Huang, and Yang \cite{Wu1959,Sawada1959}.
 The first order self-energy in Fig.~\ref{fig:diagrams}(a,b) gives the mean-field energy  shift  $\Sigma_1=\T_vn$, where ${\T}_v=2\pi a/m_{r}$ is the  boson-impurity scattering amplitude at zero energy,  with $m_r=m_{\B}m/(m_{\B}+m)$ the reduced mass. This shift is independent of temperature, and in order to get a non-trivial $T$-dependence, we need to go to second order.

The six possible second order diagrams are shown in Fig.\ \ref{fig:diagrams}. Diagrams (c-f) yield the ``Fr\"ohlich'' contribution
\begin{align}
&\Sigma_2^F(\p,\omega)=n_0(T)\T_v^2
\sum_\k\left[\frac{1}{\ek^{\B}+\ek}\right.\nonumber \\
&\left.+\frac{\ek^\B}{E_{\mathbf k}}\left(\frac{1+f_\k}{\omega - E_{\mathbf k}-\epsilon_{{\mathbf k}+{\mathbf p}}}+\frac{f_\k}{\omega + E_{\mathbf k}-\epsilon_{{\mathbf k}+{\mathbf p}}}\right)\right],
\label{Frohlich}
\end{align}
where the frequency $\omega$ is taken to have an infinitesimal positive imaginary part.
The first term 
in the integrand comes from  replacing the bare boson-impurity interaction
$g$ with the scattering matrix  $\T_v$ (see, e.g., Ref.~\cite{Christensen2015}). 
These diagrams are  non-zero only for $T\le T_c$, as they correspond to the scattering of a boson into or out of the condensate. 
The term $\Sigma_2^F$ can also be obtained from the Fr\"ohlich model~\cite{Huang2009,Novikov2009,Casteels2014}.
 
The ``bubble" diagrams (g-h) of Fig.~\ref{fig:diagrams} give
\begin{gather} \notag
\Sigma_2^B(\p,\omega) =  \T_v^2
\sum_\k
\left[v_{\mathbf k}^2(1+f_\k)\Pi_{11}(\k+\p,\omega-E_\k)\right. 
\\ 
-u_{\mathbf k}v_{\mathbf k}
[(1+f_\k)\Pi_{12}(\k+\p,\omega-E_\k)+\nn\\
\left. f_\k\Pi_{12}(\k+\p,\omega+E_\k)]
+u_{\mathbf k}^2f_\k\Pi_{11}(\k+\p,\omega+E_\k)\right]
\label{Bubble}
\end{gather}
where  the pair propagators $\Pi_{11}$ and $\Pi_{12}$ are given in Appendix \ref{pairprops}. 
The bubble diagrams have not previously been evaluated, as they require particles excited out of the condensate and consequently are  suppressed  by a factor $\sqrt{n_0\ab^3}$ for $T\ll T_c$ compared with the Fr{\"o}hlich diagrams.
Their magnitude, however, increases with $T$ as particles get thermally excited out of the BEC,  and $\Sigma_2^B$ is indeed the only non-zero contribution to second order for $T>T_c$. Note that the Fr\"ohlich model does not include
$\Sigma^B_2$ 
and therefore cannot describe  the polaron correctly for finite $T$ \cite{Christensen2015}. 

\begin{figure}[t]
\centering
\includegraphics[width=.9\columnwidth]{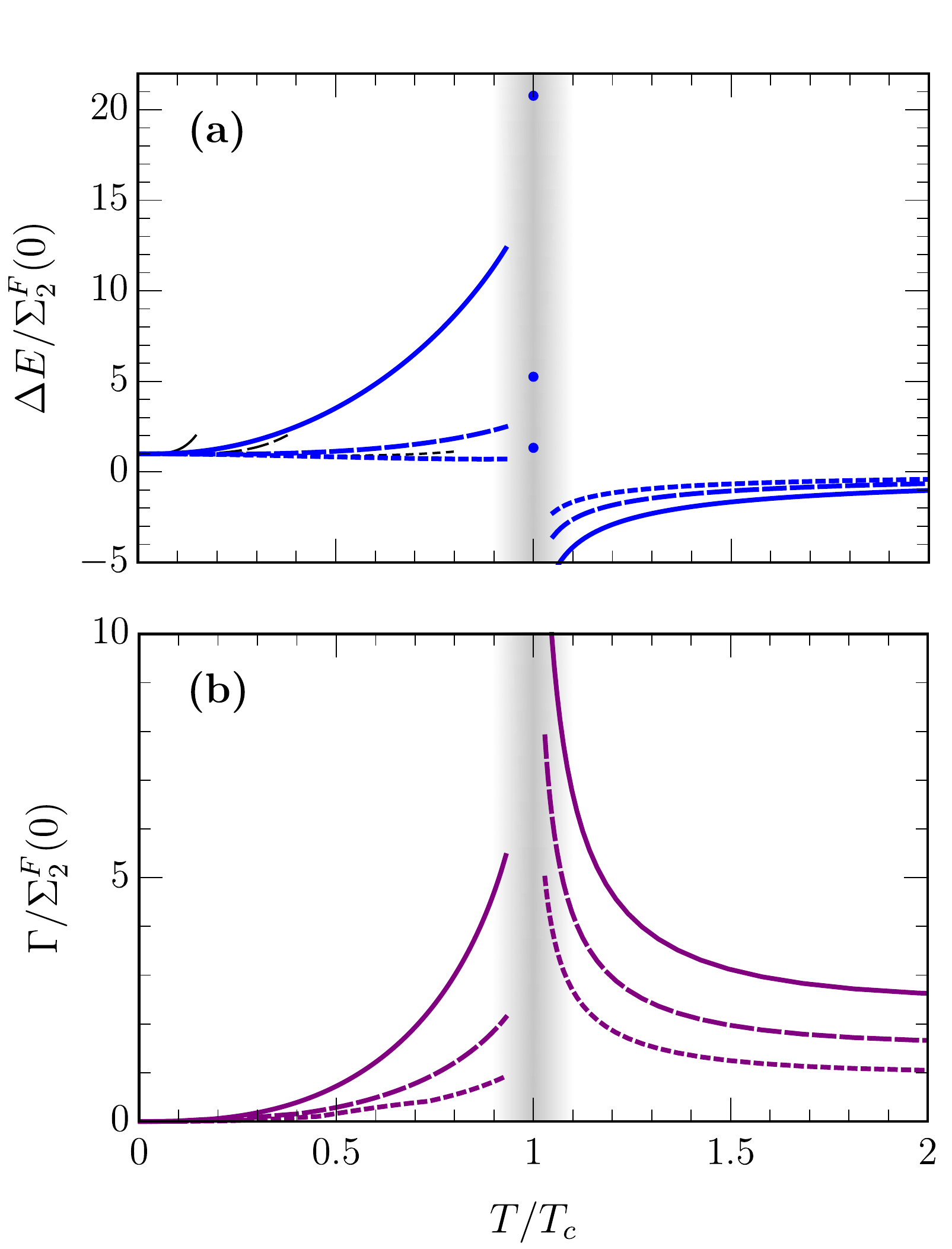}
\caption{(a) Second order energy shift and (b) decay rate for $m=\mb$. The lines are for $n^{1/3}\ab$ taking the values $0.04$ (solid), $0.1$ (dashed), and $0.25$ (short dashed). 
In (a) we also show the $T=T_c^-$ prediction \eqref{TcFrohlichFinal} for the three interaction values (dots), as well as the low-temperature prediction to fourth order in $T/T_c$ (thin, black).
The shaded region illustrates where Popov theory is expected to fail.
\label{fig:deltaE}}
\end{figure}

\section{Bose polaron at finite temperature}
\label{sec:results}

The polaron energy $E_{\p}$ for a given momentum $\p$ is  found by solving $E_{\p}=\epsilon_\p+{\text{Re}}[\Sigma(\p,E_\p)]$. Here, we focus on an  
impurity  
with momentum $\p=\0$. To second order in $a$, it is sufficient to evaluate the self-energy for zero frequency~\cite{Christensen2015},
and the equation for the polaron energy therefore simplifies to
\begin{align}
E=\text{Re}[\Sigma(\0,0)]=\T_vn+\text{Re}[\Sigma_2^F(\0,0)+\Sigma_2^B(\0,0)].
\label{PolaronEnergy}
\end{align}
The broadening of the polaron  is given by $\Gamma=-{\rm Im}[\Sigma_2^F(\0,0)+\Sigma_2^B(\0,0)]$. To simplify the notation,
we will suppress the momentum and energy arguments of the self-energy, as these are zero.
Instead, we will write $\Sigma(T)$ to focus on the $T$-dependence.

Our main results for the second-order polaron energy shift, $\Delta E\equiv E-\T_vn$, and broadening $\Gamma$
are shown in Fig.~\ref{fig:deltaE} for $m=\mb$.
We observe a strong temperature dependence, along with an intriguing non-monotonic behavior across the phase transition. We discuss the various regimes and limiting cases in the following.
For concreteness, we mainly discuss the case of equal masses $\mb=m$, with the equations for $\mb\neq m$ relegated to the appendices.

\subsection{Low-temperature behavior}
The  term $\Sigma_2^F(T)$ can be evaluated analytically for $T=0$, giving~\cite{Novikov2009,Casteels2014,Christensen2015}
\begin{align}
\Sigma_2^F(0)=\frac{32\sqrt{2}}3\frac{a^2n_0}{m\xi_0},
\end{align}
where $\xi_0$ is the healing length $\xi=1/\sqrt{8\pi n_0\ab}$ evaluated at zero temperature. An analytic expression for general mass ratio is given in Ref.~\cite{Novikov2009}. 

When evaluating $\Sigma_2^B$, we find that it contains terms that  diverge logarithmically at large momentum. This is similar
to the third order logarithmic divergence  in the polaron energy at $T=0$~\cite{Christensen2015}. 
The divergence can be cured by including the momentum dependence of the scattering matrix, which 
 provides an  ultraviolet cut-off at the  
scale $1/k=a^*\sim{\rm max}(a,\ab)$. Since the healing length sets the lower limit in the momentum integral,  we find 
\begin{align}
\Sigma_2^B(0)\simeq
\frac{4\sqrt{6\pi} a^2 n_0}{m\xi_0}
\left(\frac{2\pi}{3\sqrt{3}}-1\right)\sqrt{n_0\ab^3}\ln(a^*/\xi),
\label{eq:S2B}
\end{align}
where we ignore terms of order $(n_0 a\ab)^2$. Equation~\eqref{eq:S2B} is suppressed by $(n_0\ab^3)^{1/2}$ compared with  $\Sigma_2^F(0)$, and we thus ignore 
the terms in $\Sigma_2^B$ 
that give rise to this 
divergence 
and focus on the remainder, 
denoted 
$\tilde \Sigma_2^B(T)$ (see Appendix \ref{app:bubble} for details). 
Note that a divergent term of the form \eqref{eq:S2B} in the self-energy  is to be expected, since at $a=\ab$ the polaron ground state energy must correspond to the chemical potential of a weakly interacting Bose gas, i.e., $E=\partial E_\text{WS}/\partial n$, with $E_\text{WS}$ the energy of the weakly interacting Bose gas including the correction by Wu and Sawada \cite{Wu1959,Sawada1959}. From this argument, we also conclude that there must be a similar contribution arising from the Fr{\"o}hlich type diagrams if we treat the excitations of the BEC beyond Bogoliubov theory. Such an investigation is beyond the scope of this work.

To proceed, we take advantage of how the  self-energy below $T_c$ simplifies into a product of a $T$-dependent prefactor and a function of $\xi/\lambda$, where 
 $\lambda=(2\pi/\mb T)^{1/2}$ is the de Broglie wavelength. 
 Specifically 
\begin{align}
\Sigma_2^F(T)=\Sigma_2^F(0)\left(\frac{n_0(T)}{n_0(0)}\right)^{3/2}[1+{\cal I}_F(\xi/\lambda)].\label{eq:frohlich}
\end{align}
Here ${\cal I}_{F}$ is a   
dimensionless form of the integral appearing in (\ref{Frohlich}), see Appendix \ref{app:bubble} for details. It vanishes  at $T=0$ and its imaginary part at low temperature is only non-zero when $m<m_B$ (Appendix \ref{app:im}). Similarly to Eq.~\eqref{eq:frohlich}, an expression 
for $\tilde\Sigma_2^B(T)$  which explicitly contains the additional suppression  factor 
  $(n_0a_B^3)^{1/2}$ is given in Appendix~\ref{app:bubble}.

Due to the suppression factor, at low temperature we neglect $\tilde\Sigma_2^B$ and focus on $\Sigma_2^F$. Here, 
the superfluid density $n_0(T)$ decreases as $T^2$ for $T\ll T_c$~\cite{Glassgold1960,Shi1998}, which from Eq.~\eqref{eq:frohlich} gives a $T^2$ decrease in the polaron energy. Indeed, expanding Eq.~\eqref{eq:npopov} at low temperature yields
\begin{align}
\frac{n-n_0(T)}n \simeq  
\frac{\pi^{3/2}\left(T/T_c\right)^2}{6\zeta(\frac32)^{4/3}(n\ab^3)^{1/6}} 
-\frac{\pi^{7/2}\left(T/T_c\right)^4}{480\zeta(\frac32)^{8/3}(n\ab^3)^{5/6}},
\label{eq:popovlowT3}
\end{align}
where at each order in $T/T_c$ we keep only the leading order contribution in $n\ab^3$.
However, we find that ${\cal I}_F(\xi/\lambda)\propto(na_B^3)^{-4/3}(T/T_c)^4$ for $T\ll T_c$, and since this increase is 
proportional to $(n\ab^3)^{-4/3}$, it quickly dominates for a 
weakly interacting BEC. As a result, we obtain 
\begin{align}
E(T)\simeq E(0)+\frac{\pi^2}{60}\frac{a^2}{\ab^2}\frac{T^4}{nc^3},
\label{EFlowT}
\end{align}
where we have introduced the speed of sound in the BEC: $c=(4\pi \ab n)^{1/2}/m$. 
Interestingly, the low $T$ dependence 
of the polaron energy \eqref{EFlowT} can be related to the free energy of phonons in a weakly interacting BEC for $T\ll T_c$: $F_{\rm{ph}}=-\pi^2 T^4/(90c^3)$~\cite{Khalatnikov2000}. Indeed, setting $a=\ab$ we find that \eqref{EFlowT} exactly matches the change in the BEC chemical potential due to the thermal excitation of phonons, i.e.~$\Delta\mu=\left.-\partial F_{\rm ph}/\partial n\right|_{T,V}$. To our knowledge, this  $T^4$ increase in the chemical potential of a weakly interacting BEC has never been measured. Our result thus suggests a way to measure this effect using for instance 
radio-frequency (RF) spectroscopy on the impurity~\cite{Jorgensen2016,Hu2016}.

\subsection{Behavior close to $T_c$}\label{ClosetoTc}
We now turn our attention to temperatures close to $T_c$. From Eq.\ (\ref{Frohlich}) it follows 
that $\Sigma_2^F(T)\propto n_0(T)$ and one would at first sight expect that it vanishes as  $T\rightarrow T_c^-$. This is in fact \emph{not} the case when $m=\mb$. 
Expanding Eq.\ (\ref{Frohlich}) to lowest order in $n_0$ yields
\begin{align}
\Sigma_2^F(T_c^-)=\frac{\T_v^2}{{\mathcal T}_{\B}} 
\sum_\k f_\k=4\pi\frac{na^2}{m\ab}.
\label{TcFrohlichFinal}
\end{align}
Thus,  $\Sigma_2^F(T)$ has a \emph{non-zero} value $\propto 1/\ab$ when $T\rightarrow T_c^-$. 
Since $\Sigma_2^F$ obviously is zero for $T>T_c$, this means that it is 
\emph{discontinuous} at $T_c$. The  origin of this surprising result is that the low energy  spectrum of the Bose gas changes from linear to quadratic in momentum at $T_c$, increasing the  density-of-states dramatically. Consequently,  
the diagram given by Fig.\ \ref{fig:diagrams}(d), describing the scattering of the impurity on a thermally excited boson,
develops an infrared divergence for $n_0\rightarrow 0$ when $m=m_B$. 
For $m\neq m_B$, we on the other hand find $\Sigma_2^F(T_c^-)=0$ so that $\Sigma_2^F$ is continuous across $T_c$, see Appendix \ref{app:bubble}. 
 
Above $T_c$,    $\tilde\Sigma_2^B(T)$ is the only non-zero second-order term and 
 Eq.\ \eqref{Bubble} simplifies considerably since $v_{\mathbf k}=0$
 and $E_{\mathbf k}$ becomes $\epsilon_k^{\B}+\tb n-\mu$; i.e.\ Popov theory corresponds to 
 the Hartree-Fock approximation for $T>T_c$. The boson
  chemical potential  
  is therefore $\mu=\mu_{\text{id}}+\tb n$, 
  with  
 $\mu_\text{id}$  the chemical potential of an ideal Bose gas. We obtain 
 \begin{gather}
\hspace{-16mm}
\frac{\Sigma_2(T>T_c)}{\Sigma_2^F(T=0)} =-\frac1{\sqrt{n_0^{1/3}(0)\ab}}\left[{\cal I}_N (T/T_c)\vphantom{\left(\frac T{T_c}\right)^2}\right.
\nn \\\left. 
+i\frac{3\sqrt{\pi}[{\rm Li}_2(z)+\frac12\log^2(1-z)]}{16\zeta^{4/3}(3/2)}\left(\frac T{T_c}\right)^2\right],
 \label{eq:aboveTc}
\end{gather}
where we have used the ideal Bose gas relation  
$n\lambda^3={\rm Li}_{3/2}(z)$, with ${\rm Li}$ the polylogarithm and $z\equiv\exp(\mu_{\rm id}/T)$ the fugacity. 
The dimensionless function ${\cal I}_N (T/T_c)$ is given in Appendix \ref{app:aboveTc}. 
It follows from Eq.~\eqref{eq:aboveTc} that the imaginary part of the self-energy 
diverges  as $\log^2(1-z)$ when  $z\rightarrow 1$ for $T\rightarrow T_c^+$. 
This comes from infrared divergences in the integrals containing the 
Bose distribution function. Physically, it
means that the  polaron becomes strongly damped close to $T_c$. 
 The real part of $\Sigma_2(T)$
can also be shown 
to diverge 
when $T\rightarrow T_c^+$ as outlined in Appendix \ref{app:aboveTc}. 

\begin{figure}
\centering
\includegraphics[width=.95\columnwidth]{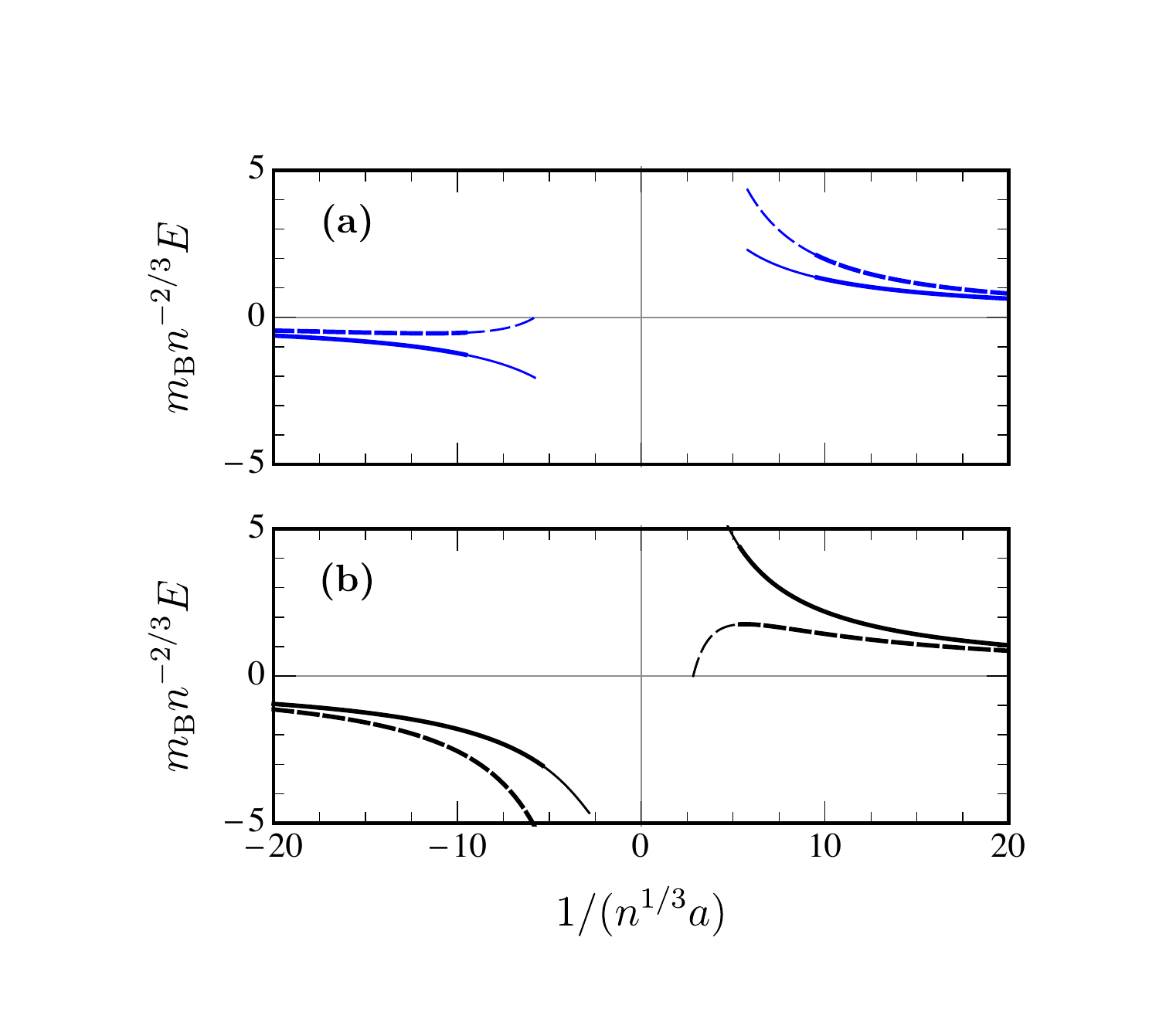}
\caption{Polaron energy as a function of interaction strength. (a) $m=\mb$ and $n^{1/3}\ab=0.003$ as in the Aarhus experiment~\cite{Jorgensen2016} for $T=0$ (solid line) and $T=T_c/10$ (dashed). (b) $m/\mb=40/87$ and $n^{1/3}\ab=0.03$ as in the JILA experiment~\cite{Hu2016} with $T=0$ (solid line) and $T=T_c/2$ (dashed). The lines are thinner in the regime $a^2>\ab\xi_0$ where the polaron ceases to be a well-defined quasiparticle~\cite{Christensen2015}, and they are only plotted in the range where the finite-temperature 2nd order shift is smaller than the mean-field energy. Note that our perturbative results are reliable at a higher temperature in the JILA experiment since the gas parameter $n^{1/3}\ab$ is larger than in the Aarhus experiment.
\label{fig:Eplot}}
\end{figure}

\subsection{High-temperature behavior}
Finally, we consider the limit $T\gg T_c$. Expanding the self-energy to lowest order in the fugacity $z$ yields 
\begin{align}
\frac{\Sigma_2^B(T)}
 {\Sigma_2^F(0)}& \simeq -\kappa
 \left[0.315 \frac{T_c}T+i\frac{3\sqrt{\pi}}{16\zeta(3/2)^{1/3}}\sqrt{\frac T{T_c}}
\right]
\label{HighT}
 \end{align}
 with $\kappa=[n_0(0)\ab^3]^{-1/6}$.
Thus, whereas the energy shift of the polaron decreases
with increasing temperature, the polaron becomes increasingly damped as the impurity collides with more and more energetic bosons. 

\subsection{Validity of perturbation theory}
At $T=0$,  the small parameter of  perturbation theory  
is $a/\xi$ and we 
additionally require $a^2/\ab\xi\ll1$ for the polaron to be 
well-defined~\cite{Christensen2015}. In general,
we expect perturbation theory to be valid provided $\Sigma_2<\Sigma_1$. From this, we derive the condition $|a|\ll\ab$ valid close to $T_c$, by 
comparing (\ref{TcFrohlichFinal}) with the first order shift $\T_vn$. For a small gas parameter, $n^{1/3}\ab$, this condition is much stricter than the $T=0$ conditions. We therefore 
expect perturbation theory to break down earlier for temperatures close to $T_c$. 
Above $T_c$, perturbation theory is accurate when 
$n^{-1/3},\lambda\gg |a|$. 
Note also that  perturbation theory  breaks 
 down  in the critical region  $|T-T_c|/T_c\lesssim n^{1/3}\ab$~\cite{Shi1998, Andersen2004}, which is the origin of the 
 infrared divergences as  
 $T\rightarrow T_c$.  However,  the critical region  is narrow for a weakly interacting BEC, making our results reliable except very close to $T_c$.

 \subsection{Numerical results}
In Fig.~\ref{fig:deltaE}, we plot the second-order self-energy $\Sigma_2$ as a function of $T$, evaluated numerically using Eq.~\eqref{eq:frohlich} for various values of the gas parameter. We see an intriguing  non-monotonic temperature dependence of  both the
polaron energy shift 
and damping. For $T< T_c$, 
the  energy shift increases and the numerical results recover our predicted $T^4$ behavior  
in Eq.~(\ref{EFlowT}) for 
$T\ll T_c$. In particular, the rate of the increase scales with $\ab^{-7/2}$ so 
that there is a strong temperature dependence  when the gas parameter of the BEC is small. The damping of the polaron, $\Gamma=-\rm{Im}\,\Sigma_2$, also increases with $T$ as more  thermally excited bosons scatter on the impurity. 
Both the energy shift and the damping vary strongly close to $T_c$. 
This reflects both the logarithmic divergences discussed above as well as the 
discontinuous jump in the Fr\"ohlich self-energy at $T_c$ 
given by Eq.\ (\ref{TcFrohlichFinal}), which is indicated by $\bullet$'s in Fig.~\ref{fig:deltaE}. 
Since perturbation theory breaks  
down close to $T_c$, we do not plot the numerical results in this region.
 For $T>T_c$, the energy shift of the polaron decreases and it vanishes as $T\to \infty$. The predicted increase in the damping rate for $T\gg T_c$ in Eq.\ (\ref{HighT}) is not visible in the range of temperatures shown in Fig.~\ref{fig:deltaE} which focuses on the phase transition region. 

In Fig.~\ref{fig:Eplot}, we plot the total polaron energy $\Sigma_1+\Sigma_2$ 
as a function of the interaction parameter $1/n^{1/3}a$  for zero and finite temperature.
We consider both the Aarhus $^{39}$K experiment and the JILA $^{40}$K-$^{87}$Rb experiment, where the latter corresponds to the case of a light impurity. 
In the region where we expect perturbation theory to be reliable, we see that the polaron energy for the equal-mass Aarhus case is shifted significantly higher by temperature, even when $T \ll T_c$. Moreover, we find a small decay rate $\Gamma \ll \Delta E$ in this regime. 
Thus, the polaron energy shift should be measurable, as we discuss below. 
On the other hand, the light impurity in the JILA case has a finite-temperature energy shift that is negative rather than positive. The reason is that --- contrary to the equal mass case ---
$\Sigma_2^F(T)$ is now continuous across $T_c$ where it goes to zero,  as discussed in Sec.\ \ref{ClosetoTc}. Its positive contribution to the polaron energy is therefore much smaller, and  the overall temperature shift becomes negative.  The decay rate $\Gamma$ on the other hand, is comparable to $|\Delta E|$ in the regime where $|\Delta E|$ is significant
 for the JILA parameters. This can be traced to the fact that $\Sigma_2^F(T)$ develops a pole and corresponding imaginary part when $m< \mb$  --- see Appendix \ref{app:im} 
for an analytic expression for ${\rm Im} \Sigma_2^F$. 
Physically the pole originates from processes where thermally excited Bogoliubov modes scatter resonantly on the polaron. These scattering are possible since the equation $\epsilon_{\mathbf k}=E_{\mathbf k}$ has 
a solution for $m<m_B$, and they lead to decay.

\section{Discussion and conclusion} 
\label{sec:conc}

The non-trivial temperature dependence of the impurity properties close to $T_c$
 is due to quite generic physics and is not limited to the specific system at hand. It originates from the change of 
 the dispersion
 from quadratic to linear at $T_c$,
  which is
  a  consequence of the $U(1)$ symmetry  breaking resulting from the formation of a condensate. This dramatically changes the low-energy density of states of the Bose gas, which impacts the excitations that couple strongly to the impurity. Thus, similar effects should occur in  other systems involving impurities coupled to a  reservoir that undergoes a phase transition where a continuous symmetry is broken. This includes  impurities in helium mixtures~\cite{BaymPethick1991book},
  conventional or high $T_c$ superconductors~\cite{Dagotto1994}, 
  magnetic systems~\cite{Kaminski2002}, and  nuclear matter~\cite{Bishop1973}.

The temperature dependence of the polaron energy  can be investigated by RF spectroscopy of $^{39}$K atoms. In these experiments, a RF pulse transfers a small fraction of atoms from a BEC in the $\ket{F=1,m_F=-1}$~state into the $\ket{1,0}$~state, such that they form  mobile impurities. The impurity-BEC interaction is highly tunable using a Feshbach resonance and thus the polaron energy can be 
obtained both for attractive and repulsive interactions. As shown in Fig.~\ref{fig:Eplot},
the energy shift due to a finite temperature is sizable in the regime where perturbation theory should be reasonable: at $1/(n^{1/3}a)=10$ the energy at  $T=T_c/10$ compared to $T=0$ corresponds to a RF frequency shift of $\sim7$~kHz, which is comparable to the experimental resolution.
 Since the temperature dependence of the polaron energy scales with $\Sigma_2^F(0)\propto a^2 n_0$, it is favorable to access a given interaction strength by choosing a large scattering length and accordingly small density.   
 
To conclude, using perturbation theory valid in the weak coupling regime, we investigated the properties of the Bose polaron as a function of temperature. We derived 
analytical results both for low temperature $T\ll T_c$,  $T\simeq T_c$, and high temperature $T\gg T_c$.
These results show that the superfluid phase transition of the surrounding Bose gas has strong effects on the properties of the polaron. The energy depends in a non-trivial way on $T$ with a  pronounced non-monotonic behaviour around $T_c$, and  the damping 
increases sharply as $T_c$ is approached. We argued that these effects should occur in a wide range of systems consisting of impurities  immersed in an environment 
undergoing a phase transition. Finally, we discussed how 
this intriguing temperature dependence can be detected experimentally.

\acknowledgements
We thank M.~W.~Zwierlein for pointing out the interesting analogy between  Eq.~(\ref{EFlowT}) and the energy of a phonon gas in a BEC. 
 We appreciate useful discussions with B.~Zhu.
JL, MMP, and GMB acknowledge financial support
from the Australian Research Council via Discovery Project
No.~DP160102739. JL is supported through the Australian
Research Council Future Fellowship FT160100244. 
JL and MMP acknowledge funding from the Universities Australia -- Germany Joint Research Co-operation Scheme.
GMB wishes to acknowledge the support of the Villum Foundation via grant VKR023163. 
JA acknowledges support from the Danish Council for Independent Research
and the Villum Foundation.
This work was performed in part at the Aspen Center for Physics, which is supported by the National Science Foundation Grant No.~PHY-1607611.




\onecolumngrid

\appendix

\section{Pair propagators}
\label{pairprops}
After performing the Matsubara frequency sums, we obtain 
\begin{align}
      \Pi_{11}(p) =\sum_\k
  \left[\frac{ u_\k^2(1+f_{\k})}{z - E_\k-\epsilon_{\k+\p}}+\frac{ v_\k^2f_{\k}}
    {z + E_\k-\epsilon_{\k+\p}} + \frac{2 m_r}{k^{2}}\right]
\label{PairPropagator}
\end{align}
for the normal pair propagator at four-momentum $p=(\p,z)$, and
\begin{align}
  \Pi_{12}(p) &=-T
  \sum_{\omega_\nu}G_{12}(-\k,-\omega_\nu)G(\k+\p,\omega_\nu+z)=\sum_\k
  \left[\frac{u_\k v_\k(1+f_{\k})}{E_\k+\epsilon_{\k+\p}-z}
    +\frac{u_\k v_\k f_{\k}}{\epsilon_{\k+\p} - E_\k- z}\right],
\label{AnomalousPropagator}
 \\ 
  \Pi_{22}(p) &=-T
  \sum_{\omega_\nu}G_{22}(-\k,-\omega_\nu)G(\k+\p,\omega_\nu+z)=\sum_\k
  \left[\frac{ u_\k^2f_{\k}}{z + E_\k-\epsilon_{\k+\p}}+
    \frac{v_\k^2(1+f_{\k})}{z - E_\k-\epsilon_{\k+\p} }\right]
\label{ParticleholePropagator}
\end{align}
for the anomalous and particle-hole propagators.

\section{Self-energy below $T_c$: Fr{\"o}hlich and bubble diagram integrals}
\label{app:bubble}

To find the polaron energy within perturbation theory, we evaluate the Fr{\"o}hlich diagrams at zero momentum and frequency, but finite temperature:
\begin{align}
\Sigma_2^F(T)&=\,n_0(T)\T_v^2
\sum_\k\left[\frac{1}{\ek+ \ek^{\B}}+\frac{\ek^\B}{E_{\mathbf k}}\left(\frac{1+f_\k}{- E_{\mathbf k}-\epsilon_{{\mathbf k}}}+\frac{f_\k}{E_{\mathbf k}-\epsilon_{{\mathbf k}}}\right)\right] \nn \\
&=\underbrace{\frac{2\pi n_0(0) a^2}{m_r\xi_0}A(\alpha)}_{\Sigma_2^F(T=0)}
\left(\frac{n_0(T)}{n_0(0)}\right)^{3/2}
\bigg[1+\underbrace{\frac2\pi\frac{1+1/\alpha}{A(\alpha)}
\int \,\bar f_\k \frac{k^2dk}{\sqrt{k^2+2}}
\left(
\frac{-1}{\sqrt{k^2+2}+k/\alpha}+
\frac{1}{\sqrt{k^2+2}-k/\alpha}\right)}_{{\cal I}_F(\xi/\lambda,\alpha)}\bigg]
\, ,
\label{eq:Frohlich}
\end{align}
where we have switched to dimensionless variables in the second line, measuring momentum in units of the inverse healing length. Here $\alpha\equiv m/\mb$ is the mass ratio and $\xi_0$ is the BEC healing length evaluated at $T=0$. The Bose distribution in dimensionless units is
\begin{align}
\bar f_\k=\frac1{\exp\left[\frac{\lambda^2}{4\pi\xi^2}k\sqrt{k^2+2}\right]-1}.
\end{align}
The function defined in the main text for equal masses is ${\cal I}_F(\xi/\lambda)\equiv {\cal I}_F(\xi/\lambda,1)$.
For 
 $T\ll T_c$ we have
\begin{equation}
{\cal I}_F(\xi/\lambda)\simeq 
\frac{\pi^4}{1280\zeta(\frac32)^{8/3}(n\ab^3)^{4/3}}\left(\frac {T}{T_c}\right)^4. 
\label{IFlowT}
\end{equation}
The mass-ratio dependent function $A$ was found for general mass ratio in Ref.~\cite{Casteels2014} (see also Ref.~\cite{Christensen2015}) to be
\begin{align}
A(\alpha)=\frac{2\sqrt{2}}{\pi}\frac1{1-\alpha}\left[1-\frac{2\alpha^2}{1+\alpha}\sqrt{\frac{\alpha+1}{\alpha-1}}\arctan\sqrt{\frac{\alpha-1}{\alpha+1}}\right],
\end{align}
with the definition $\sqrt{-1}=i$. The function $A$ is well-defined for equal masses, where
\begin{align}
A(1)=\frac{8\sqrt{2}}{3\pi},
\end{align}
which leads to $\Sigma_2^F(0)=32\sqrt{2} a^2n_0/(3m\xi_0)$.

Similarly to the  Fr{\"o}hlich diagrams, we evaluate the ``bubble" contribution. We note that again there is a contribution which is present even at $T=0$. Specifically, this is the term which does not contain a Bose distribution in any of the momentum summations. This term, however, arises from bosons excited out of the condensate (see Fig.~\ref{fig:diagrams}), and is thus suppressed by a factor $\sqrt{n_0\ab^3}$ compared with the Fr{\"o}hlich diagrams. This suppression only increases at finite temperature and therefore we ignore this term in the following. Instead, using Eq.~\eqref{Bubble} we define
\begin{align}
\tilde \Sigma_2^B(T) = & \T_v^2 
\sum_\k
\bigg\{ f_\k
[v_{\mathbf k}^2\Pi_{11}(\k,-E_\k) 
+u_{\mathbf k}^2\Pi_{11}(\k,E_\k)
-u_{\mathbf k}v_{\mathbf k}\Pi_{12}(\k,E_\k) 
-u_{\mathbf k}v_{\mathbf k}\Pi_{12}(\k,-E_\k)]
\nn  \\ & \hspace{17mm}
+ v_{\mathbf k}^2\tilde\Pi_{11}(\k,-E_\k) 
-u_{\mathbf k}v_{\mathbf k}\tilde\Pi_{12}(\k,-E_\k)
\bigg\}
\nn \\ 
= & \Sigma_2^F(T=0)\sqrt{n_0(0)\ab^3}\left(\frac{n_0(T)}{n_0(0)}\right)^2 
\frac{(1+1/\alpha)(8\pi)^{5/2}}{2A(\alpha)}\nn \\ &\times
\int 
\frac{d^3kd^3p}{(2\pi)^6}\left\{-\bar f_\k \bar f_\p\left[
\frac{\bar v_\k^2\bar u_\p^2+\bar u_\k\bar v_\k\bar u_\p\bar v_\p}{\bar E_\k+\bar E_\p+\bar \epsilon_{\k+\p}}
+\frac{\bar v_\k^2\bar v_\p^2+\bar u_\k^2\bar u_\p^2+2\bar u_\k\bar v_\k\bar u_\p\bar v_\p}{ -\bar E_\k+\bar E_\p+\bar \epsilon_{\k+\p}}
+\frac{\bar u_\k^2\bar v_\p^2+\bar u_\k\bar v_\k\bar u_\p\bar v_\p}{-\bar E_\k-\bar E_\p+\bar \epsilon_{\k+\p}}
\right]
\right.\nn \\ & \hspace{22mm}\nn \left.
-\bar f_\k\left[
\frac{\bar u_\k^2\bar v_\p^2+\bar v_\k^2\bar u_\p^2+2\bar u_\k\bar v_\k\bar u_\p\bar v_\p}{\bar E_\k+\bar E_\p+\bar \epsilon_{\k+\p}}
+\frac{\bar v_\k^2\bar v_\p^2+\bar u_\k^2\bar u_\p^2+2\bar u_\k\bar v_\k\bar u_\p\bar v_\p}{-\bar E_\k+\bar E_\p+\bar \epsilon_{\k+\p}}
-\frac{\bar u_\k^2+\bar v_\k^2}{(1+\alpha)\bar \epsilon_\p}
\right]\right\}
\\ 
\equiv & \Sigma_2^F(T=0)\sqrt{n_0(0)\ab^3}
\left(\frac{n_0(T)}{n_0(0)}\right)^2{\cal I}_B(\xi/\lambda,\alpha).
\label{eq:Bubble}
\end{align}
Here, we made the integral in the first line dimensionless by extracting a factor $2m_B/\xi^4$, and defining the dimensionless functions $\bar E_\k=k\sqrt{2+k^2}$, $\bar \epsilon_\k=k^2/\alpha$, $\bar u_\k=\sqrt{\frac{k^2+1}{2E_\k}+\frac12}$ and $\bar v_\k=\sqrt{\frac{k^2+1}{2E_\k}-\frac12}$.
$\tilde \Pi_{ij}$ refers to the pair propagator including only those terms involving the Bose distribution function $f_\k$, as we ignore the term which 
is suppressed at zero temperature (see discussion in the main text). Comparing Eq.~(\ref{eq:Bubble}) with Eq.~(\ref{eq:Frohlich}) explicitly shows that it is  suppressed by a  
factor $(n_0a_B^3)^{1/2}$.

The bubble diagrams contain several simple poles, which we treat numerically by introducing a small imaginary part, i.e., by taking $z\to z+i\delta$ (in the above, this can be achieved by shifting $\bar\epsilon_\k$ slightly below the real axis), and then extrapolating our results to $\delta=0$. We estimate the relative error in the evaluation of the bubble diagrams resulting from this procedure to remain well below $1\%$ for all $\xi/\lambda$ considered.

\begin{figure}
\centering
\includegraphics[height=.32\columnwidth]{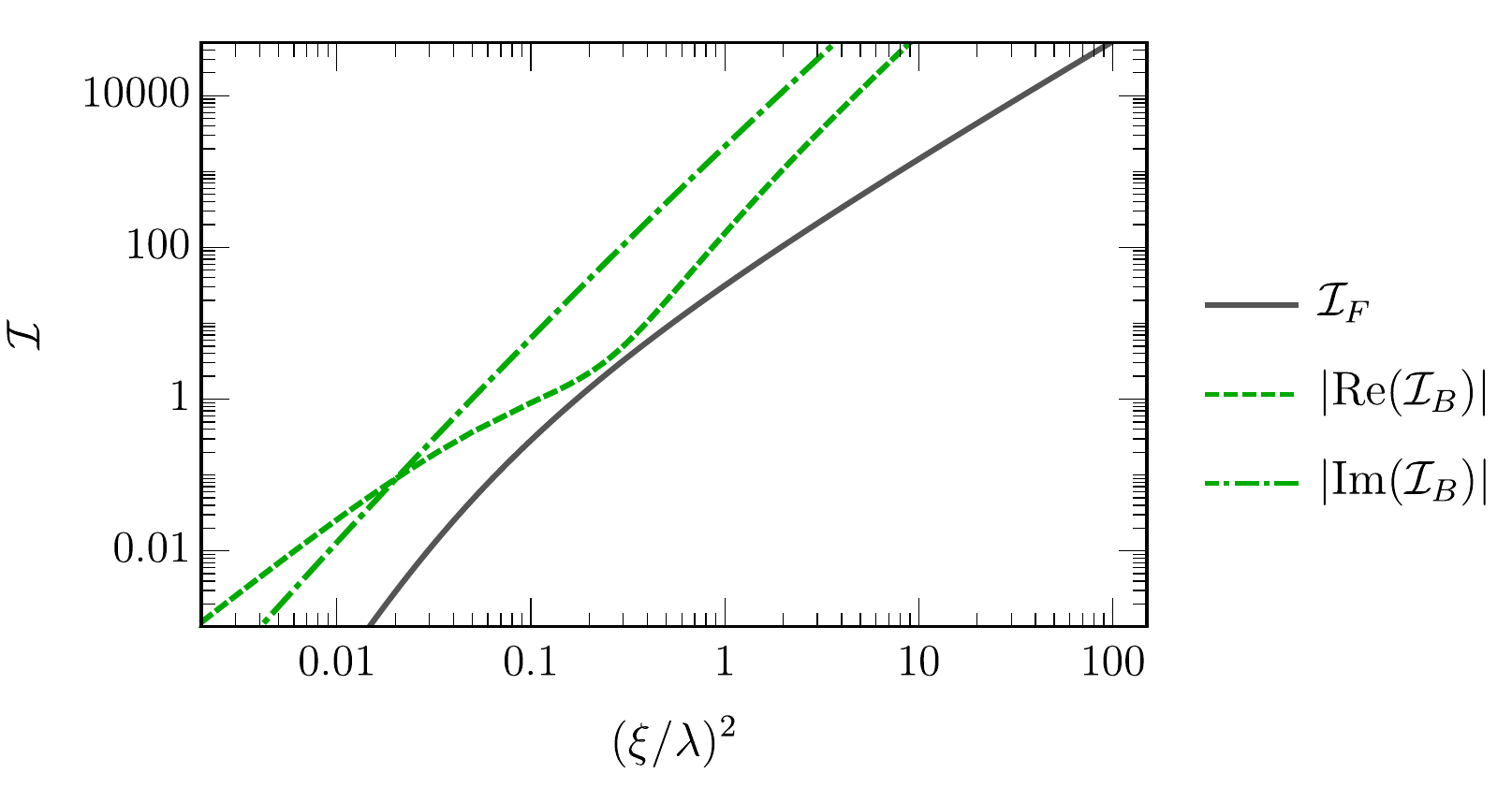}
\caption{The functions ${\cal I}_F$ (solid, black) and the real (green, dashed) and imaginary (green, dot-dashed) parts of ${\cal I}_B$, calculated for equal masses $\mb=m$. The latter two are negative, therefore we take the absolute values of these. 
\label{fig:integrals}}
\end{figure}

In Fig.~\ref{fig:integrals} we show the result for the dimensionless functions ${\cal I}_F$ and  ${\cal I}_B$ for equal masses. In this case, the Fr{\"o}hlich diagrams are purely real, and we see that they are larger than the bubble diagrams except at very small (outside the range shown) or large temperature.

\begin{figure}
\centering
\includegraphics[height=.35\columnwidth]{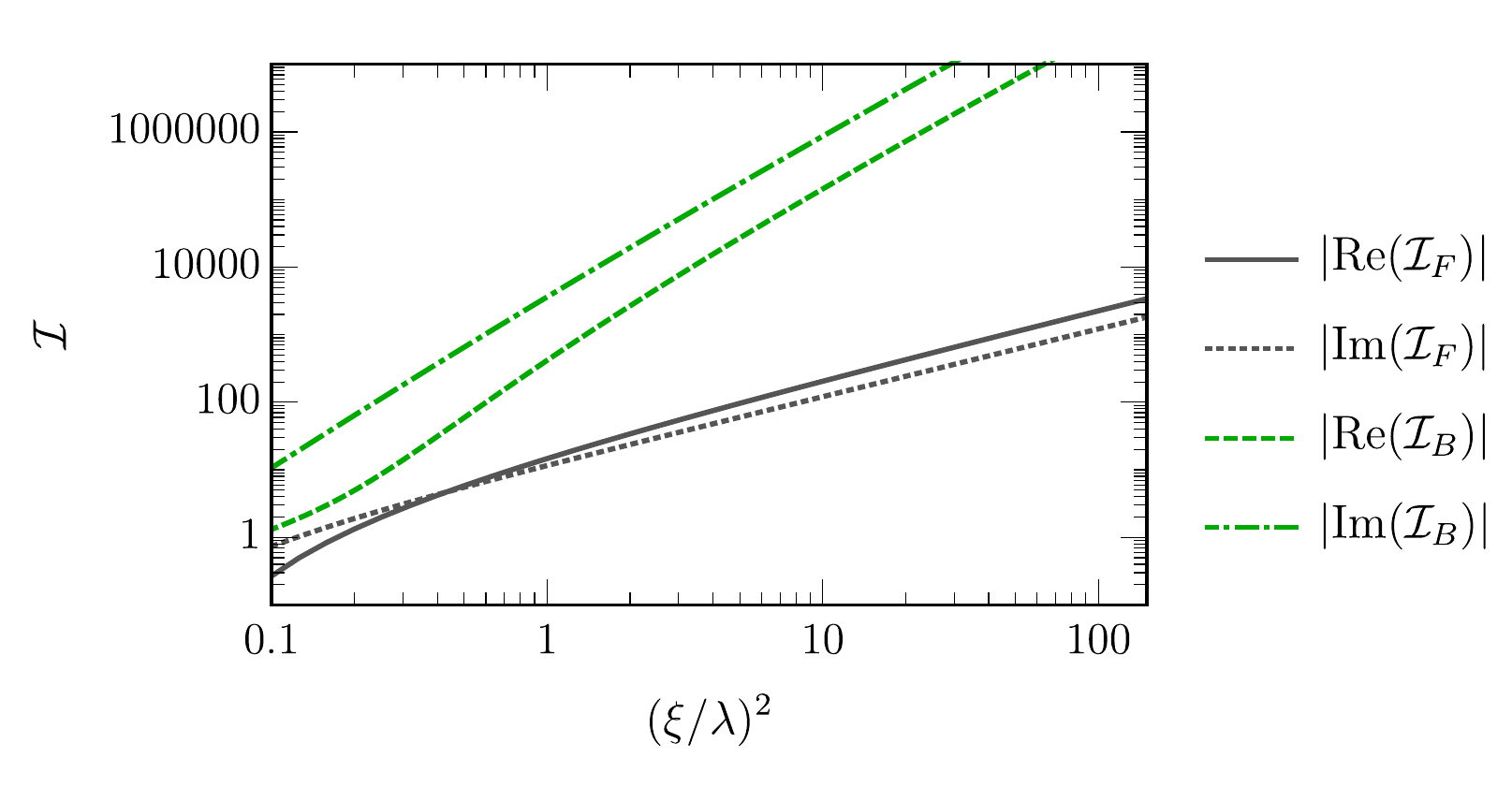}
\caption{The dimensionless integrals for a $^{40}$K impurity immersed in a $^{87}$Rb BEC. We show the real (black, solid) and imaginary (black, dotted) parts of ${\cal I}_F$ together with the real (green, dashed) and imaginary (green, dot-dashed) parts of ${\cal I}_B$. These are all negative within the range shown. 
\label{fig:integrals2}}
\end{figure}

For unequal masses, the main qualitative difference is that the Fr{\"o}hlich diagram develops a simple pole when $\mb>m$. This is easily integrated over, and in Fig.~\ref{fig:integrals2} we show the resulting functions  for the particular case of a $^{40}$K atom immersed in a $^{87}$Rb condensate.

\section{Imaginary part of the Fr\"ohlich self-energy}
\label{app:im}

The imaginary part of the Fr\"ohlich self-energy for zero momentum and frequency is found from Eq.~(\ref{Frohlich}) of the main text to be
\begin{align}
{\rm Im}\Sigma_2^F(T)=-n_0(T)\T_v^2
\sum_\k\frac{\ek^\B}{E_{\mathbf k}}f_\k\delta( \epsilon_{\mathbf k}- E_{\mathbf k}).
\label{ImFrohlich}
\end{align}
It follows that the imaginary part is non-zero only if $\epsilon_{\mathbf k}=E_{\mathbf k}$ has a solution, i.e.\ if $m<m_B$. Doing the integral (\ref{ImFrohlich}) yields 
\begin{align}
{\rm Im}\Sigma_2^F(T)=-\frac2\pi\frac{\alpha^3}{(1-\alpha^2)^{3/2}} \T_v^2 n_0(T)^{3/2}m_B^{3/2}\T_B^{1/2}f_{\mathbf k_0}
\label{ImFrohlichFinal}
\end{align}
where $\mathbf k_0$ is the $\mathbf k$ vector which solves $\epsilon_{\mathbf k}=E_{\mathbf k}$.

\section{Self-energy above $T_c$}
\label{app:aboveTc}

Above $T_c$, the second order self energy reduces to the term from the bubble diagrams
\begin{align}
\Sigma_2(T>T_c) & ={\cal T}_v^2\sum_\k f_\k\Pi_{11}(\k,E_\k)=
{\cal T}_v^2\sum_\k f_\k\sum_\p \left(\frac{1+f_\p}{\ek^\text{B}-\ep^\text{B}-\epsilon_{\k-\p}+i0}+\frac1{\ep^\text{B}+\ep}\right) \nn \\
& =
{\cal T}_v^2 8\mb^2m_rT^2\sum_\k \frac1{e^{k^2}/z_\text{id}-1}\sum_\p \left[
\left(\frac1{p^2}-\frac1{p^2-k^2/\gamma^2-i0}\right)+
\frac1{e^{(\p+\k/(1+\alpha))^2}/z_\text{id}-1}\frac1{k^2/\gamma^2- p^2+i0}\right],
\label{eq:sigmalargeT}
\end{align}
where in the second line we shifted $\p\to\p+\k/(1+\alpha)$ in all terms except the renormalization (last term of the first line). We also measured momenta in units of $\sqrt{2\mb T}$ and defined the ratio $\mb/m_r\equiv\gamma$. The quantity $z_\text{id}\equiv e^{\mu_\text{id}/T}$ is the fugacity of the ideal Bose gas. It is related to the density through
\begin{align}
n\lambda^{3}=\lambda^{3}\sum_\k f_\k=\mbox{Li}_{3/2}(z_\text{id}),
\label{eq:naboveTc}
\end{align}
and can be further related to $T/T_c$ through the ideal gas expression
\begin{align}
T/T_c=[\zeta(3/2)]^{2/3}(n\lambda^3)^{-2/3}.
\label{eq:naboveTc2}
\end{align}

\begin{figure}[t]
\centering
\includegraphics[height=.35\columnwidth]{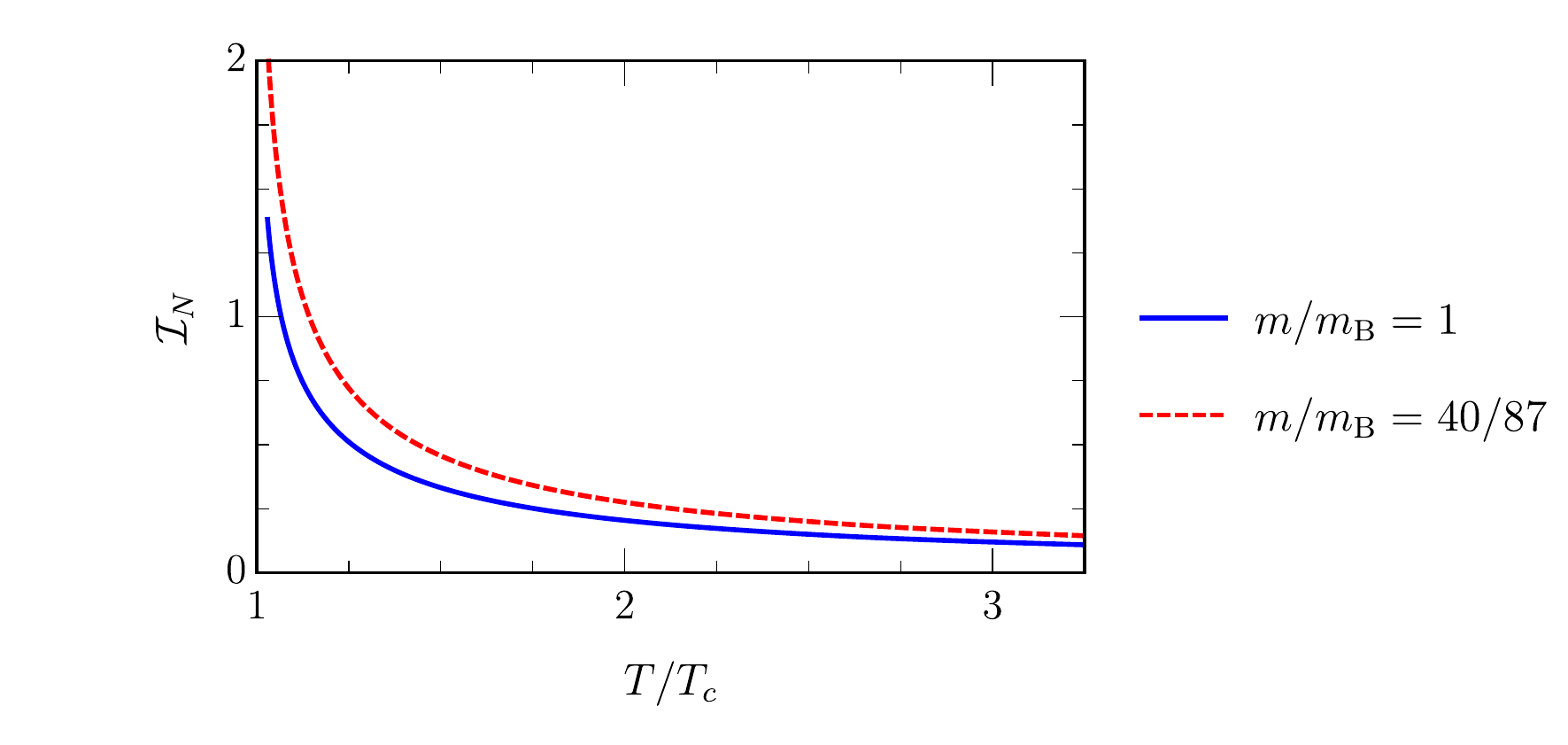}
\caption{The dimensionless integral appearing in the self-energy for temperatures above $T_c$. We show the result both for equal masses (blue, solid), and for a $^{40}$K impurity immersed in a $^{87}$Rb BEC (red, dashed).
\label{fig:Inc}}
\end{figure}

To proceed, we note that the integral over the angle between $\k$ and $\p$ in Eq.~\eqref{eq:sigmalargeT} can be performed analytically:
\begin{align}
\int_{-1}^1dx\frac1{e^{a+bx}-1}=\frac1b\log\frac{e^a-e^{-b}}{e^a-e^b},
\label{eq:trick}
\end{align}
assuming $a>b>0$. Since the integral over the term in parenthesis in the second line of Eq.~\eqref{eq:sigmalargeT} is purely imaginary, we have
\begin{align}
\frac{\mbox{Re}[\Sigma_2(T>T_c)]}
{\Sigma_2^F(T=0)}
=
&\frac{{\cal T}_v^2 
4\mb^2mT^2}{\Sigma_2^F(T=0)}\frac1{8\pi^4}\int_0^\infty 
\frac{k\,dk}{e^{k^2-\mu_b/T}-1}\int_0^\infty \frac{p\,dp\, 
{\cal P}}{k^2/\gamma^2-p^2}
\log \frac{e^{p^2+k^2/(1+\alpha)^2-\mu_b/T}-e^{-2kp/(1+\alpha)}}{e^{p^2+k^2/(1+\alpha)^2-\mu_b/T}-e^{2kp/(1+\alpha)}}
\nn \\
 &\hspace{-20mm}=\frac{-1}{\sqrt{n_0(0)^{1/3}\ab}}\underbrace{\sqrt{\frac2{\pi^3}}\frac{1+\alpha}{A(\alpha)}\frac{(T/T_c)^2}{\zeta^{4/3}(3/2)}
\int_0^\infty \frac{k\,dk}{e^{k^2-\mu_b/T}-1}\int_0^\infty \frac{p\,dp\, {\cal P}}{p^2-k^2/\gamma^2}
\log \frac{e^{p^2+k^2/(1+\alpha)^2-\mu_b/T}-e^{-2kp/(1+\alpha)}}{e^{p^2+k^2/(1+\alpha)^2-\mu_b/T}-e^{2kp/(1+\alpha)}}}_{{\cal I}_N(T/T_c,\alpha)},
\label{eq:reS}
\end{align}
where ${\cal P}$ indicates that only the principal part of the integral should be evaluated. The prefactor which scales as $1/\sqrt{\ab}$ arises from the normalization by $\Sigma_2^F(T=0)$. The integral ${\cal I}_N $ is evaluated numerically, and the result is shown in Fig.~\ref{fig:Inc}. The function referenced in the main text Eq.~\eqref{eq:aboveTc} is ${\cal I}_N (T/T_c)\equiv{\cal I}_N (T/T_c,1)$.

For equal masses, the imaginary part of the self-energy can be determined analytically for all $T > T_c$:
\begin{align}
\frac{\mbox{Im}[\Sigma_2(T>T_c)]}
{\Sigma_2^F(T=0)}  = 
- 
\frac{3\sqrt{\pi}}{16\zeta^{4/3}(3/2)}\frac1{\sqrt{n_0(0)^{1/3}\ab}}\left(\frac T{T_c}\right)^2
\left[{\rm
 Li}_2(z) +\frac{1}{2} 
 \log^2(1-z) 
 \right] 
\end{align}
For a $^{40}$K impurity in a $^{87}$Rb condensate, we evaluate the imaginary part of the self energy numerically, again using Eq.~\eqref{eq:trick}.

\subsection{Logarithmic divergence of $\Sigma_2^B$ above $T_c$}

For concreteness, we focus on equal masses. One can  rewrite the integral appearing in Eq.~\eqref{eq:reS} at $T_c$ as:
\begin{align}
\sum_{\k,\p} \frac{f_\k f_\p}{\ek-\ep-\epsilon_{\k-\p}+i0} & = - \frac{2 T^2 m^3}{(2\pi)^4} \int dp \int dk \frac{p k}{\left(e^{k^2/2} -1 \right)\left(e^{p^2/2} -1 \right)} \log\left[\frac{p+k -i0}{p-k-i0}\right] \\
& \simeq \frac{T^2 m^3}{\pi^4}\int dr \frac{1}{r} \int d\phi \frac{1}{\sin\phi \cos\phi} \log\left[\frac{\cos\phi+\sin\phi -i0}{\cos\phi-\sin\phi-i0}\right] \\
& \to -(19.71-21.78 i) \frac{T^2 m^3}{\pi^4} \log(r).
\end{align}
Here we have made the transformation $p = r \cos\phi$, $k = r \sin\phi$, and then considered the regime $r\ll 1$. Thus, we see that this integral diverges logarithmically as $r \to 0$.

\twocolumngrid

%

\end{document}